\documentclass[twocolumn,showpacs,preprintnumbers,amsmath,amssymb,superscriptaddress]{revtex4-1}

\usepackage{graphicx}
\usepackage{graphics}
\usepackage{dcolumn}
\usepackage{bm}
\usepackage{color}
\usepackage{soul}

\begin{document}

\title{Radiolysis of water confined in porous silica: A simulation study of the physicochemical yields}

\author{H. Ouerdane}
\affiliation{CIMAP, unit\'e mixte CEA-CNRS-ENSICAEN-UCBN 6252 BP 5133, F-14070 Caen, Cedex 05, France}
\author{B. Gervais}
\affiliation{CIMAP, unit\'e mixte CEA-CNRS-ENSICAEN-UCBN 6252 BP 5133, F-14070 Caen, Cedex 05, France}
\author{H. Zhou}
\affiliation{LIRIS, IPNL, UMR IN2P3-CNRS-UCBL 5822, Villeurbanne, F-69622, France}
\author{M. Beuve}
\affiliation{LIRIS, IPNL, UMR IN2P3-CNRS-UCBL 5822, Villeurbanne, F-69622, France} 
\author{J.-Ph. Renault}
\affiliation{CEA/Saclay, DSM/IRAMIS/SCM/URA 331 CNRS, 91191 Gif-sur-YVette Cedex, France}

\date{\today}

\begin{abstract}
We investigate the radiolysis of liquid water confined in a porous silica matrix by means of an event-by-event Monte Carlo simulation of electron penetration in this composite system. We focus on the physical and physicochemical effects that take place in the picosecond range, before the radicals start to diffuse and react. We determine the radiolytic yields of the primary species for a system made of cylindrical pores filled with water over a wide range of pore radii $R_{\rm C}$. We show that the relative position of the conduction band edge $V_0$ in both materials plays a major role in the radiolysis of composite systems. Due to its lower $V_0$ as compared to that of silica, water acts as a collector of low-energy electrons, which leads to a huge enhancement of the solvated electron yields for $R_{\rm C} \leq$ 100 nm. The confinement has also a marked effect on the spatial distribution of the radicals, which become isolated in a very large number of pores as $R_{\rm C}$ decreases.
\end{abstract}


\maketitle
\section{Introduction} 
The radiolysis of water can be drastically perturbed when it takes place in confining media and more generally in situations where interfaces represent a significant fraction of the volume.
A remarkable example is that of water deposited on ZrO$_2$ surfaces, for which the yield of H$_2$ formation can reach extraordinarily large values \cite{ref1}. Many other materials are known to modify the radiolytic yields in solution, as reported in recent experimental works \cite{ref2}. In the case of silica nanometric particles in water solution, an increase of H$_2$ yields was observed \cite{ref3,ref4}. It was suggested that this excess of H$_2$ production could arise from an excess of electrons originally created in silica, which cross the solid-liquid interface \cite{ref3}. Porous silica was also used as a confining material. In such a case, the geometry is reversed and water is confined in nanometric pores. The radiolysis of water confined in silica at a nanometric scale shows a diminution of hydroxyl yields as the pore size decreases \cite{ref5}. Enhanced radiolytic yields were also reported in this case for H$_2$ and H$_2$O$_2$.\cite{ref6} Confinement effects on radiolytic yields were also observed in micellar systems \cite{ref7}. 

Our understanding of the various phenomena at work in confined water radiolysis is far from complete, and theoretical works or simulations like those existing for pure liquid water \cite{ref8,ref9,ref9b,ref10} are scarce. Many aspects regarding the characteristics of such a heterogeneous system can influence the water radiolysis. They can be roughly divided into two parts. First, the so-called physical and physicochemical stages may be modified with respect to pure homogeneous water. These stages include the detail of energy release in the composite system, the following short-time electron thermalization and hole diffusion, possibly leading to water/silica interface crossings, the quick rearrangement of the excited or ionized water molecules, and eventually the formation of long-lived transient species, like excitons, in the solid part of the material. All these processes can be affected by the geometry of the material and, of course, by the nature of the solid phase. Second, the chemical stage, which results from the diffusion and reaction of the radicals generated during the early stage in water, may be strongly perturbed by the presence of a solid phase. The most obvious perturbation is the limitation of the diffusion process by the solid/liquid interface, which depends on the geometry of the composite system. Some other questions arise regarding the role of the interface itself, where some specific reactions can take place, like the decomposition of excitons initially created in silica or catalytic reactions. It is thus worth seeking methods that allow us to determine the radiolytic yields in composite systems, taking into account their composite nature and their particular geometry, in order to identify the key parameters that control the generation of radicals in water. 

To this end, we present a detailed event-by-event simulation, which aims at describing the radiolysis of water confined in porous silica. The interaction of radiations with pure liquid water and bulk silica has been studied for several decades, and our knowledge of the necessary input to perform a simulation is sufficient to achieve reasonably accurate predictions. We focus here on the so-called physical and physicochemical stages, which cover typically the first picoseconds following the interaction of radiation with matter. Taking into account the electronic structures of both materials, the physical stage describes the interaction of the projectile with the composite system in terms of electronic excitation, ionization, attachment, thermalization, and interface effects on the subpicosecond time scale. It thus provides the characteristics of the electronic cascade resulting from ionizing events. The physicochemical stage is dominated by proton transfer in liquid water and by electron-hole recombination in silica. It provides a three-dimensional spatial distribution of all species and the corresponding physicochemical yields. This is the starting point of the subsequent chemical stage, which accounts for the diffusion of molecular products and radicals, over a time scale ranging from a few picoseconds to microseconds and beyond. The simulation of the chemical stage will be the object of future work. 

Addressing the problem of radiation interaction with matter in its whole complexity would be a very cumbersome task, and some simplifying assumptions are necessary. From a geometrical
point of view, the actual vycor glasses often used to confine water are rather complex. For the sake of simplicity, we consider a regular array of cylindrical pores of radius $R_C$ embedded in amorphous silica and filled with liquid water. The interface is assumed to be perfectly defined, and we consider that each part of the material can be described from its bulk properties alone. From these two hypotheses, it is possible to build a Monte Carlo simulation of electron transport through the composite system and to investigate the role of its geometric structure on a nanometer scale.

The simulation provides the physicochemical yields which are of paramount importance to interpret experimental results. In this paper we simulate the impact of a 50 keV electron in
the system. It is representative of a piece of ionization cascade generated by more energetic light particles, either electrons or high-energy photons, whose tracks have a comparable three-dimensional structure. The exact energy does not really matter, provided that the range of the electron is large with respect to the typical length scale of the material. We obtain the radiolytic yields for a series of pore radius $R_{\rm C}$ ranging from 1 to 10$^3$ nm. We discuss the sensitivity of our results with respect to several parameters like the electron-phonon interaction cross sections and the porosity $p$, defined as the ratio of the water volume over the whole volume of the sample. We end our discussion with an analysis of the segregation, defined as the ability of the system to isolate a limited number of radicals from each other in different pores. 

\section{Simulation}

\subsection{Geometrical aspects and materials properties} 
We consider as a model a periodic distribution of cylindrical pores over the whole sample. For convenience, we divide the sample into square cells of side length $L$, which characterizes the in-plane periodicity of the model system. Each cell contains one cylinder. For the cylindrical geometry investigated here, the porosity of the system reads $p = R_{\rm C}^2/L^2$. 
This simplified geometry is expected to give a reasonable description of confinement on water radiolysis, while avoiding an unnecessary complication of the simulation code. It is possible to build more realistic geometrical models of water in porous silica, which would account, for example, for connections between randomly distributed pores of various sizes. However, it would complicate substantially the numerical treatment of the electron transport for little gain in terms of the description of the physicochemical effects of nanometric confinement on water radiolysis. 

\begin{figure}
	\centering
		\includegraphics[width=0.50\textwidth]{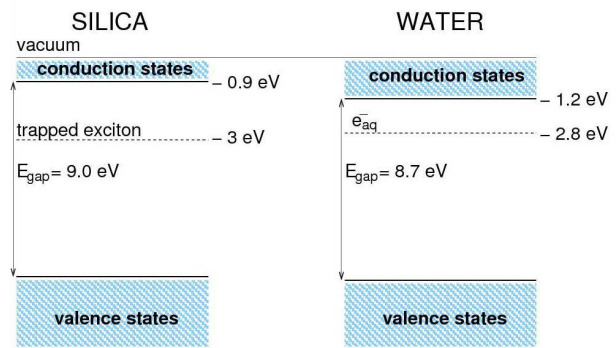}
	\caption{Schematic representation of the energy levels in water and silica. Note that the tops of the valence bands are fortuitously at the same energy in silica and water.}
	\label{fig1}
\end{figure}

Much information about the electronic structure of each material, such as the electronic density of states \cite{ref15} and optical energy loss functions \cite{ref16,ref17} is available. From these data, it is possible to derive collision cross sections and energy loss probabilities, as discussed in the next section. To describe the transport of electron through a composite material, we need however to locate the electronic levels in both material with respect to each other. The exact positions of the conduction band edges with respect to vacuum level, $V_0$, are not accurately known for both materials. We use $V_0 = -1.2$ eV for water \cite{ref11} and $V_0=-0.9$ eV for silica \cite{ref12}. The latter value depends actually on several parameters like silica band gap and Si affinity used as reference. From the data of Ref. \cite{ref12} and for a band gap varying between 9.0 eV \cite{ref13} and 9.3 eV \cite{ref14} the value of $V_0$ is found between $-$0.2 and $-$0.9 eV. For water, the value of $V_0$ might actually be closer to the vacuum level as well. As we shall see in this paper, this is the difference between these values, $\Delta U = V_{0,{\rm SiO}_2} - V_{0,{\rm H}_2{\rm O}}$, which is of critical importance rather than the absolute values themselves. At the interface between water and silica, the excited electrons experience therefore a potential step $\Delta U$. Other specific effects related to the interface, like surface states, polarization, structural disorder, or the presence of chemical species such as silanols, are neglected in the present study.

The difference between the valence and conduction band edges is taken to be 9.0 eV in silica \cite{ref12,ref13,ref14} and 8.7 eV in water \cite{ref11} as depicted in Fig. \ref{fig1}. The specific mass is 1.0 g/cm$^3$ for water and 2.25 g/cm$^3$ for silica. 

\subsection{Electron and hole transport} 
\begin{figure}
	\centering
		\includegraphics[width=0.50\textwidth]{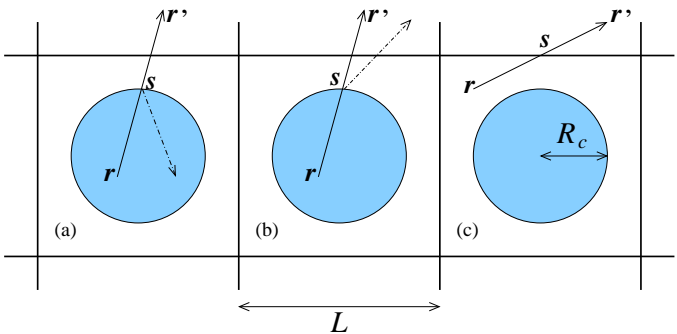}
	\caption{Schematic representation of transport simulation at interface: (a) reflection at interface; (b) transmission through interface; (c) virtual surface crossing.}
	\label{fig2}
\end{figure}

During the course of its transport, an electron can travel through both silica and water. The energy of an electron in a given medium is taken as $E=p^2/2m^{\star} + V_0$, where $V_0$ and $m^{\star}$ are the corresponding conduction band edge and effective mass, respectively, and ${\bm p}$ is the electron momentum. For silica, the effective mass is taken from Ref. \cite{ref18}. For water, we simply assume that the effective mass is equal to the free electron mass. The transport is simulated by a series of elastic and inelastic collisions in the composite system. Each collisional event is separated from the next one by a time of free flight $t$ determined from a Poisson law characterized by its mean time-of-flight $\tau$. According to the location of the electron, either in water or in silica, a free flight is sampled and a new position ${\bm r}'$ is generated from the previous position of the electron ${\bm r}$, its velocity ${\bm v}= {\bm p}/m^{\star}$, and the time-of-flight $t$: ${\bm r}' = {\bm r} + {\bm v}t$. If the new position is outside of the initial cell or cylinder, the point of interface crossing ${\bm s}$ is determined, as depicted in Fig. \ref{fig2}. At the interface between two cells, the medium does not actually change (virtual interface, case (c) in Fig. \ref{fig2}). The new position is reset to this point, i.e., ${\bm r}' = {\bm s} + \delta {\bm e}_{\bm v}$, where $\delta$ is a tiny positive number which ensures that the electron effectively changes cell and ${\bm e}_{\bm v}$ is the unit vector along the velocity vector ${\bm v}$. If a real interface is encountered (cases (a) and (b) in Fig. \ref{fig2}), the potential energy changes by an amount of $\Delta U$, and the momentum of the electron needs to be changed accordingly.

The components of momentum after interaction with the interface potential step are obtained from the conservation of energy: 

\begin{equation}
\frac{{p_{||}'}^2}{2m^{\star '}} = \frac{p_{||}^2}{2m^{\star}} ~~~~ \mbox{and} ~~~~ \frac{{p_{\perp}'}^2}{2m^{\star '}} = \frac{p_{\perp}^2}{2m^{\star}} + \Delta U
\end{equation}

\noindent where ${\bm p}_{||,\perp}$ are the components of ${\bm p}$ parallel and perpendicular to the surface, respectively. When $p_{\perp}^2/2m^{\star} + \Delta U < 0$, the electron does not have enough momentum to overcome the surface potential step. It is thus specularly reflected and ${\bm p}' = {\bm p}_{||} - {\bm p}_{\perp}$. The electron is then located on the surface at ${\bm r}' = {\bm s} + \delta {\bm e}_{\bm v}$ when it crosses the interface [case (b)] or ${\bm r}' = {\bm s} - \delta {\bm e}_{\bm v}$ when it is reflected [case (a)]. For a system made of water and silica, the latter case occurs only when the electron is initially in water. After interaction with the interface, a new free flight is sampled, according to the medium in which the electron is located. 

In our simulation, we follow all the electrons until their kinetic energy becomes lower than 37 meV, i.e., lower than the average thermal energy $3k_{\rm B}T/2$ at $T = 300$ K, where $k_{\rm B}$ is the Boltzmann constant. We assume that the thermalized electrons either recombine with the surrounding holes or get solvated on spot, when they are located in water. We also take into account the limited hole transport by sampling the migration distances according to a Gaussian distribution whose mean value is given by the migration length $d$ in the medium considered. For water, $d=1.4$ nm can be deduced from the experiment of Ogura and Hamil \cite{ref19}. For silica, the diffusion of pre-thermalized holes before trapping is believed to be smaller by 1 or 2 orders of magnitude than electron diffusion \cite{ref24}. For migration in the subpicosecond range investigated here, we tentatively assume that the characteristic distance is of the order of 1 oxygen-oxygen interatomic distance and take $d=0.2$ nm. For the sake of completeness, we also consider hole migration across the interface. As it can be observed in Fig. \ref{fig1}, this is energetically possible. However, little is known regarding this process. We will discuss the significance of this hypothesis with the results. 

\subsection{Mean free path}
\begin{figure}
	\centering
		\includegraphics[width=0.50\textwidth]{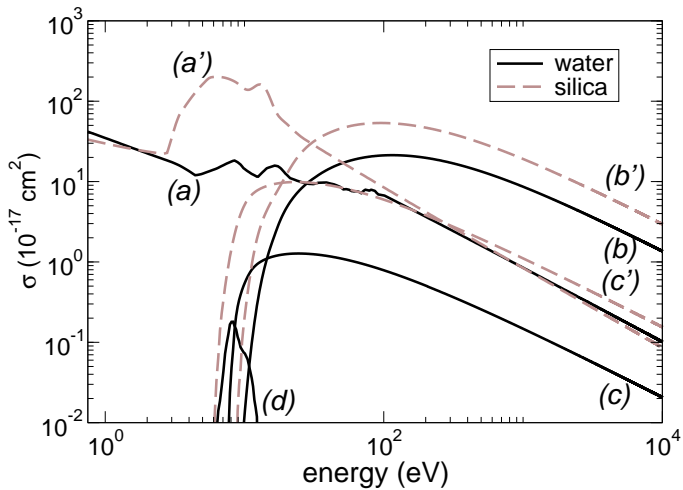}
	\caption{Electron interaction cross sections for water (continuous lines, a, b, c, d) and silica (long-dashed lines, a', b', c'): (a, a') vibrational excitation; (b, b') ionization; (c, c') excitation; (d) electron attachment to water molecule.}
	\label{fig3}
\end{figure}

In this section, we restrict the description of the physical stage to its main aspects. Technical details related to electron transport can be found in Ref. \cite{ref10} and in the Appendix. The data for water were discussed by several authors \cite{ref9,ref10,ref20} and we put here more emphasis on silica. The mean time-of-flight $\tau$ is related to the mean free path $\lambda$ and to the cross section $\sigma$ by the relation $\lambda^{-1} = \tau^{-1}/v = n\sigma$, where $n$ is the molecular density of the medium in which the event takes place and $v$ is the electron velocity. The cross sections we use in our simulation are represented in Fig. \ref{fig3}. 

For both media, the ionizing collisions responsible for electron multiplication dominate significantly all of the other processes for electron energy greater than a few tens of electronvolts. Our cross sections are based on the work of Kim and Rudd \cite{ref21}. We have adapted this formulation to take into account the densities of states of the valence and conduction bands. The details of the parametrization are given in the Appendix. The inner-shell ionization processes have relatively low cross sections. However, these rare events lead to large energy losses, which contribute significantly to the stopping power. For light elements like Si or O, these processes are followed by Auger electron emission while radiative recombination is negligible. We take into account the whole Auger cascade, which generates electrons at an energy of several hundreds of electronvolts, according to the ionized inner shell. Such a process leads to a relatively dense sequence of ionizing events. Our parametrization gives an average energy for electron-hole pair creation $W_{{\rm SiO}_2}$ of the order of 19.1 eV, in agreement with other empirical estimates \cite{ref22}. For comparison, $W_{{\rm H}_2{\rm O}} \approx 20.7$ eV. These values correspond to the raw electron-hole pair creation, before recombination is taken into account. The value of $W_{{\rm SiO}_2}$ is sensitive to the relative proportion of ionization and excitation. However, varying by 20\% one of these contributions does not change significantly this value.

When the electron kinetic energy becomes of the order of a few tens of electronvolts, the electronic excitation becomes the most efficient energy loss process. Following previous works
for water \cite{ref9,ref10}, we build a model cross section from the optical oscillator strength. For silica, we consider one single excitation channel leading to the formation of an exciton. The corresponding oscillator strength is deduced from the energy loss function of $\alpha$-quartz \cite{ref17}. The corresponding parametrization is given in the Appendix.

When the electron kinetic energy is smaller or comparable to the band gap, vibrational excitation of the medium becomes overwhelming. The cross sections for emission and absorption
of longitudinal optical (LO) phonons and for acoustic phonons have been extensively studied for silica in Refs. \cite{ref26,ref27,ref28,ref29}. For acoustic phonons, the cross section was fitted to experimental results \cite{ref32} in the energy range 8-18 eV and extrapolated continuously above 18 eV according to Ref. \cite{ref18}. We proceeded in the same fashion for water. We used the vibrational excitation cross section as measured by Sanche et al. \cite{ref23,ref23b} and previously used for bulk water radiolysis \cite{ref10,ref20}. In addition, electron attachment to water molecules is made possible for energies ranging from 6.25 to 12.4 eV. The procedures used to obtain the energy-dependent cross sections for these processes are presented in detail in Ref. \cite{ref10}. 

As observed by Cartier and co-workers for silica, the electron-phonon interaction becomes extremely efficient for kinetic energies between 4 and 20 eV, and a quick slowing down takes place in this energy range \cite{ref32}. Below the energy threshold for LO-phonon emission, interaction with acoustic phonon is the only possible process. Since phonon absorption and emission have comparable probabilities in our simulation, the thermalization sometimes requires a very large number of collisions. Therefore, the distributions of thermalization times exhibit a long tail extending up to several picoseconds. For silica, the maximum of these distributions ranges from 25 to 150 fs in the energy range 1-8 eV, while the corresponding average thermalization times range from 250 to 350 fs. These simulated values are consistent with experimental observations of fast transient optical processes \cite{ref30}. The thermalization times in water are somewhat comparable \cite{ref20}. During thermalization, the electrons can be transported over several nanometers in both media. In pure silica, for example, at an energy of 3 eV, the distribution of thermalization distances peaks at 8 nm and the average value is roughly 30 nm. 

\subsection{Electron-hole recombination}
The physical stage ends when all electrons are thermalized, typically a few hundreds of femtoseconds after the projectile impact. Following thermalization, several processes start competing with each other. For water, the large dielectric constant allows electrons to solvate, but a fraction of them may recombine with holes, before ionized water molecules dissociate. During the same time the holes can migrate. In silica, owing to a low dielectric constant, the physicochemical stage is governed by the long-range Coulomb interaction between electrons and holes. The spatial distribution of electrons and holes in the composite system is inhomogeneous, and the precise description of the diffusion-recombination stage requires in principle to follow the correlated motion of all electrons and holes. 

The most salient feature of this stage can nevertheless be obtained by considering that all pairs are independent of each other. Under this assumption, the probability $P(r)$ of recombination for a pair made of one electron and one hole separated from each other by a distance $r$ is given by \cite{ref25}

\begin{equation}\label{eq2}
P(r) = \frac{R_{\rm O}}{r_{\rm eff}}~{\rm erfc} \left(\gamma~\frac{r_{\rm eff} - R_{\rm O}}{R_{\rm O}}\right)
\end{equation}

\noindent where $r_{\rm eff} = R_{\rm O}\left[1- \exp(-R_{\rm O}/r)\right]^{-1}$ and $R_{\rm O} = q^2/(4\pi\epsilon_0\epsilon_{\rm r}k_{\rm B}T)$ is the Onsager radius, which depends on the relative dielectric constant $\epsilon_{\rm r}$ of each medium and on the temperature $T$; and $q$ the electronic charge. The constant $\gamma = R_{\rm O}/(4Dt)^{1/2}$ is the characteristic rate of electron-cation recombination, which depends on the time $t_{\rm rec}$ allowed for recombination. We use $t_{\rm rec}=$ 1 ps in our simulation. The values of the diffusivity $D$ can be estimated from the electronic mobility $\mu$, via the Einstein relation $D=\mu k_{\rm B}T/q$. This value corresponds to electrons accelerated in an electric field, for which the kinetic energies are larger than the thermal energy. Alternatively, it can be estimated from the mean free path $\lambda$ and the mean time of free flight $\tau$ by identifying the mean-squared displacement length deduced from the Poisson distribution of mean free path to the mean-squared displacement deduced from a Gaussian distribution associated with Brownian motion. Such an approach gives the relation $\lambda^2=3D\tau$. As expected from the energy dependence of the mean free path, this second estimate gives a larger value, but the orders of magnitude of both estimates are comparable. The values of the parameters are given in Table \ref{tab1}. The larger value of the Onsager radius in silica reflects the lower dielectric constant of this material. 

\begin{table}[!rh]
\centering
\caption{Values of $\epsilon_{\rm r}$, $R_{\rm O}$, $D$, and $\gamma$ in water and silica at room temperature.}
\label{tab1}
\begin{tabular}{cccccc}
\hline\hline
~&~&~&~&~&\\
~&~&$\epsilon_{\rm r}$&$R_{\rm O}$ (nm)&$D$ (nm$^2$s$^{-1}$)&$\gamma$\\
~&~&~&~&~&\\
H$_2$O&~&78 &0.72&1.4$\times 10^{13}$&0.2\\
~&~&~&~&~&\\
SiO$_2$&~&3.9 &14.0&0.5--3.0$\times 10^{14}$&0.8\\
~&~&~&~&~&\\
\hline\hline
\end{tabular}
\end{table}

Our simulation of electron-hole pair recombination is very similar to the independent reaction time, often used in the simulation of inhomogeneous chemical reaction in water radiolysis \cite{ref31} We sample a probability of recombination according to Eq. \eqref{eq2}. The allowed recombinations are then sorted by decreasing order of probability. If a particle is involved in more than one recombination, only the most likely recombination is retained for this particle. For the sake of simplicity, the recombination process is simulated separately in each medium. This hypothesis will be discussed with the results in the next sections. 

\section{Results and discussion}

Our simulation provides the yields, defined as the number of a given species generated in water divided by the energy deposited in the whole sample. We focus on the three main species at the end of the physicochemical stage: aqueous electron $e_{\rm aq}^-$, hydroxyl radical OH, and hydronium ion H$_3$O$^+$. The yields are expressed in units of 10$^{-7}$ mol/J, except when otherwise stated. 

For each pore size, we sample randomly 300 impacts of a 50 keV electron. The probability for an impact to be located in a given medium m (H$_2$O or SiO$_2$) depends on the electron generation process. It is proportional to the cross section $\sigma_{\rm m}$ for the corresponding medium and to the number of elementary structural units, either H$_2$O or SiO$_2$, which constitute the medium m. The proportion of impacts in water thus reads: 

\begin{equation}\label{eq3}
f_{{\rm H}_2{\rm O}} = \frac{n_{{\rm H}_2{\rm O}} \sigma_{{\rm H}_2{\rm O}}}{n_{{\rm H}_2{\rm O}} \sigma_{{\rm H}_2{\rm O}} + n_{{\rm SiO}_2} \sigma_{{\rm SiO}_2}}
\end{equation} 

\noindent with $n_{{\rm H}_2{\rm O}} = pV{\mathcal N}_{\rm A}\rho_{{\rm H}_2{\rm O}}/M_{{\rm H}_2{\rm O}}$, and where $V$ is the volume of the sample, $pV$ is the volume of water in the sample, ${\mathcal N}_{\rm A}$ is the Avogadro number, $\rho_{{\rm H}_2{\rm O}}$ is the specific mass of water, and $M_{{\rm H}_2{\rm O}}$ is its molar mass. We can express the above fraction with respect to the porosity $p$: 

\begin{equation}\label{eq4}
f_{{\rm H}_2{\rm O}} = \frac{p}{p + (1-p)R_{\lambda}}
\end{equation} 

\noindent where $R_{\lambda} = \lambda^{-1}_{{\rm SiO}_2}/\lambda^{-1}_{{\rm H}_2{\rm O}}$, and $\lambda^{-1}_{\rm m}$ is the inverse mean free path for the generation process in medium m. For 50 keV X-rays, the ratio deduced from the mass attenuation coefficients is $R_{\lambda}\approx 3.15$. For high-energy electrons, we obtained a ratio $R_{\lambda}\approx 1.50$, which depends weakly on the energy. A purely geometric distribution of impact is obtained for $R_{\lambda}=1.0$. For all results presented below, we used $R_{\lambda}= 1.50$. For small pore sizes, the range of a 50 keV electron is much larger than the radius of a pore, and the exact location of the impact point is unimportant. However, in the limit of large pore sizes, sampling the impacts according to Eq.~\eqref{eq4} allows one to link the yield in the composite system $Y$ to its counterpart for bulk water $\tilde{Y}$. Indeed, from the definition of the yields: $Y= \tilde{Y}E_{{\rm H}_2{\rm O}}/(E_{{\rm H}_2{\rm O}} + E_{{\rm SiO}_2})$, where $E_{{\rm H}_2{\rm O}}$ and $E_{{\rm SiO}_2}$ are the energies released in H$_2$O and SiO$_2$, respectively. 

In the limit of infinitely large pores, the proportion of energy released in a given medium is simply the energy of the radiation multiplied by the fraction of impacts in this medium, and thus
$Y =\tilde{Y}f_{{\rm H}_2{\rm O}}$. In the limit of very small pores, it is customary to consider that the dose deposited in each material is equal to the average dose. In such a case, we obtain a relationship similar to Eq. \eqref{eq4}, but with $R_{\lambda}$ substituted by $R_M=\rho_{{\rm SiO}_2}/\rho_{{\rm H}_2{\rm O}}$ or by the corresponding ratio of electronic density $R_{\rm e}$. For silica and water, $R_M=2.25$ and $R_{\rm e}= 2.0$. In both cases the dependence of $f_{{\rm H}_2{\rm O}}$ with the porosity is similar, but we prefer to use $R_{\lambda}$, which is exact in the limit of very large $R_{\rm C}$. 

\subsection{Physicochemical yields} 
\begin{figure}
	\centering
		\includegraphics[width=0.50\textwidth]{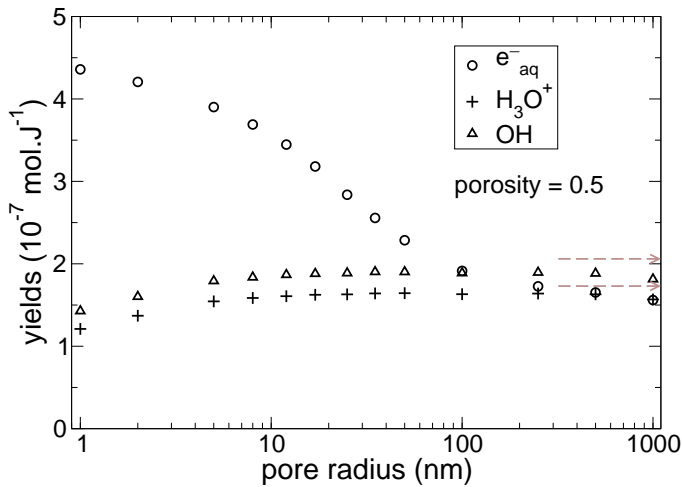}
	\caption{Yields of aqueous electrons, $e_{\rm aq}^-$, hydroxyl, OH, and hydronium, H3O$^+$, at the end of the physicochemical stage. The porosity is 0.5. The arrows on the right-hand side indicate the asymptotic limit extrapolated from bulk water values for the corresponding porosity.}
	\label{fig4}
\end{figure}

In Fig. \ref{fig4}, we present the physicochemical yields versus pore radius $R_{\rm C}$. Appreciable variations can be observed for $R_{\rm C} \leq 100$ nm. Above this value, the yields no longer vary with $R_{\rm C}$ and the formation of radicals is dominated by bulk water processes. From our simulation of bulk liquid water, the values of the yields are $\tilde{Y}_{{\rm e}_{\rm aq}^-} = 4.31$, $\tilde{Y}_{\rm OH} = 5.12$, and $\tilde{Y}_{{\rm H}_3{\rm O}^+} = 4.43$. For a porosity $p=0.5$, $f_{{\rm H}_2{\rm O}}= 0.4$ and we can deduce the following asymptotic values for $R_{\rm C} \rightarrow \infty$: $Y_{{\rm e}_{\rm aq}^-}=1.73$, $Y_{\rm OH} = 2.06$, and $Y_{{\rm H}_3{\rm O}^+} = 1.78$. These values are indicated by arrows in Fig. \ref{fig4}. At $R_{\rm C}=10^3$ nm, the simulated yields are lower than the corresponding asymptotic values by 10\% typically. This effect originates from the long-range structure of the energy deposition for a 50 keV electron. At this energy, the range of an electron is larger than the characteristic length scale of the system $R_{\rm C}=10^3$ nm, and a larger amount of energy is deposited in the denser medium, i.e., in silica in our case. The asymptotic limit is reached for $R_{\rm C} \approx 10^6$ nm.

Below 100 nm, we observe a dramatic increase of the electron yield as $R_{\rm C}$ decreases. It originates from the energy difference between the conduction band edges in water and silica, as
illustrated in Fig. \ref{fig1}. The low-energy electrons diffusing through silica that reach the interface can cross it and lose energy in water by coupling with the vibrational modes of the latter medium. Once they have lost some energy, they are no longer able to overcome the potential energy barrier of the order of 0.3 eV to climb up to the conduction band of silica. They are thus trapped in water, where they thermalize and further get solvated. This effect is rather insensitive to the exact value of the potential energy barrier $\Delta U$. Changing $\Delta U$ by a factor 4 leads to nearly identical values for the yields. However, when this energy difference becomes lower than the thermal energy, this process is modified. It is of course completely suppressed, when the energy difference becomes negative. In this case, the electrons thermalize in silica to recombine with holes, leading eventually to exciton formation. The preferential trapping in water depends exclusively on the relative position of the bottom of the conduction bands $\Delta U$ and could in principle be observed for any material with a sufficiently high $\Delta U$. 

Below $R_{\rm C}=2$ nm, the electron yield does not vary significantly and reaches a value $Y_{{\rm e}_{\rm aq}^-} = 4.4$. In this limit, a large fraction of electrons created in silica is collected in water. In order to check the reliability of this result, we shifted arbitrarily the energy loss spectrum toward lower energies by 2.5 eV. As a result, we obtained $W_{{\rm SiO}_2} = 16.4$ eV and $Y_{{\rm e}_{\rm aq}^-}=5.5$, which means that a larger number of electrons is produced in silica and thus collected in water. Conversely, shifting empirically the spectrum to obtain $W_{{\rm SiO}_2} =25$ eV, we obtained $Y_{{\rm e}_{\rm aq}^-}=3.9$ in the limit of small $R_{\rm C}$, for a porosity of 0.5. This means that the effect of electron collection in water is still clearly visible in this case. 

The trapping effect observed for electrons does not exist for the other radiolytic products. The large difference between $Y_{{\rm e}_{\rm aq}^-}$ and and $Y_{{\rm H}_3{\rm O}^+}$ for small pore radius shows clearly that the collection of electrons in water is due to subexcitation electron transport and not due to an excess of ionization in water. The large difference between electron and hydronium yields observed in Fig. \ref{fig4} implies that the a net negative charge is created in water. Conversely, the same amount of positive charge is accumulated in silica.

\begin{figure}
	\centering
		\includegraphics[width=0.50\textwidth]{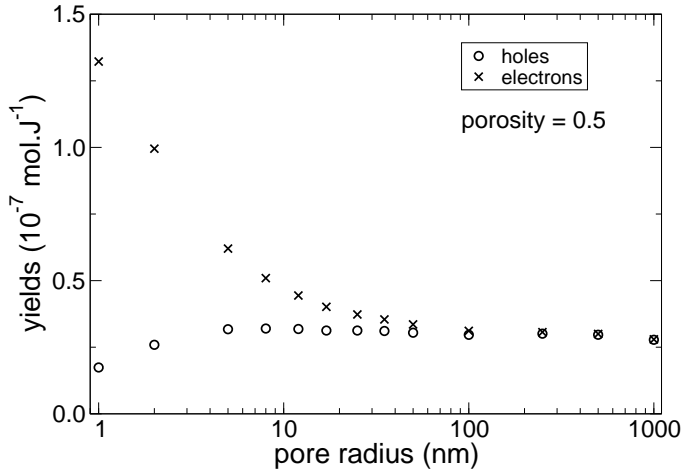}
	\caption{Yields of electron and hole loss from water. The yield of electron loss (crosses) corresponds to electron-hole pair recombination in water. The yield of hole loss (circles) corresponds to both electron-hole pair recombination and hole diffusion across the interface. The porosity is 0.5.}
	\label{fig5}
\end{figure}

Both OH and H$_3$O$^+$ species are created only in water where they originate mainly from the same ionization process, so that their yields are strongly correlated. An additional amount of OH comes from dissociative excitation and, to some extent, from electron-hole pair recombination, making $Y_{\rm OH}$ larger than $Y_{{\rm H}_3{\rm O}^+}$ by a value almost independent of $R_{\rm C}$. Since there is no additional sources for these products in pure silica, the corresponding yields remain rather constant down to $R_{\rm C} = 12$ nm. Below this value, another process contributes to reduce these yields. The holes created in water indeed migrate before the dissociation of an H$_2$O$^+$ molecule takes place \cite{ref19}. The corresponding migration length is significantly larger than the hole migration length in silica. The decrease of $Y_{\rm OH}$ and $Y_{{\rm H}_3{\rm O}^+}$ at very low $R_{\rm C}$ simply reflects that the net flux of holes at the interface is directed from water to silica. This effect can be traced back to the yield of hole loss by diffusion or recombination and to the yield of electron-hole pair recombination as shown in Fig. \ref{fig5}.

The observed yield of outgoing holes shown in Fig. \ref{fig5} is probably an upper limit. Indeed, it is assumed here that the holes diffuse freely across the interface. Little is known regarding
this process, but it is possible that the efficiency of the crossing is less than 1, and some of the holes would be either reflected or trapped at the surface. In such cases, the yield of outgoing
holes would be smaller. We checked that our result would not be drastically affected if interface crossing were strictly forbidden. In such a case, the dependency of $Y_{\rm OH}$ and $Y_{{\rm H}_3{\rm O}^+}$ on $R_{\rm C}$ almost vanishes. It is important to note that the yield of electrons in water is quite insensitive to our hypothesis regarding the hole flux at the interface.

When free interface crossing is assumed, we observe in Fig. \ref{fig5} that the net loss of holes from water lowers the recombination yield for $R_{\rm C} < 5$ nm. This small effect reflects the competition between the electron-hole pair recombination and outgoing hole diffusion, which depletes the hole population inside a given pore. Moreover, for small $R_{\rm C}$, the number of pores containing at most one electron becomes larger, as it will be discussed below. These isolated electrons have no partner available to recombine with in their respective pores, so that the probability of electron recombination decreases.

\subsection{Electron-phonon coupling strength} 
Like for any transport simulation, the quality of the results depends on the quality of the cross sections used as input parameters. It is therefore necessary to investigate the sensitivity of our results to these parameters. We study here the effects of changing the interaction cross section in silica on the yields in water. These cross sections control the propagation length of electrons and hence their ability to reach the interface between water and silica. For interface crossing, we observed that the most sensitive part along an electron trajectory is the low-energy part. When the excitation energy of an electron becomes lower than 10 eV, it loses its energy mainly by interaction with LO phonon. When its energy is low enough, it interacts mainly with acoustic phonon and hence progressively thermalizes, either in silica or in water, according to its random diffusion through the composite system. Increasing the electron-phonon interaction reduces the electron mean free path and thus reduces the diffusion length. Conversely, reducing the electron-phonon interaction increases the diffusion length. 

\begin{figure}
	\centering
		\includegraphics[width=0.50\textwidth]{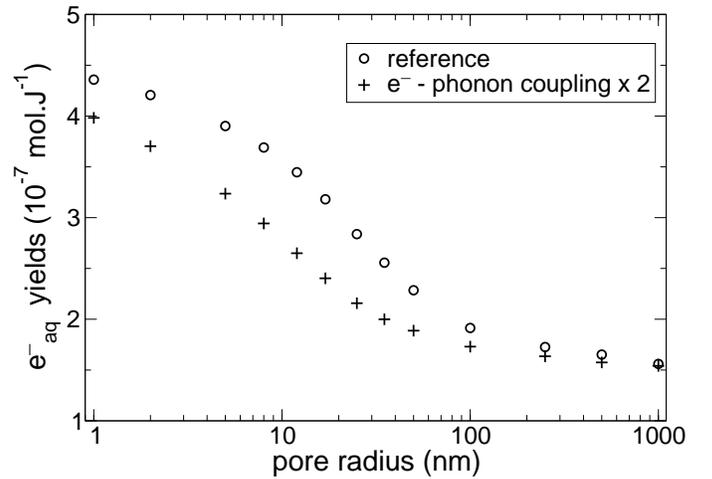}
	\caption{Yields of aqueous electrons, $e_{\rm aq}^-$, for two $e^-$-phonon interaction coupling strengths. The porosity is $p=0.5$.}
	\label{fig6}
\end{figure} 

The influence of this parameter is shown in Fig. \ref{fig6} where we compare the results obtained by assuming a twofold increase of the electron-phonon coupling strength with respect to the reference results, obtained without any scaling. For larger electron-phonon coupling, we observe that the collection of electron in water is less efficient and limited to smaller values of $R_{\rm C}$. Nevertheless, the increase of $Y_{{\rm e}_{\rm aq}^-}$ for decreasing $R_{\rm C}$ is clearly observable in both cases. This gives further confidence in our simulation work, though the electron-phonon coupling in porous silica is not known accurately. Our results show that electron-phonon coupling strength compatible with experiment on electron transport \cite{ref32}. The electron-phonon coupling in silica has of course no effect on $Y_{\rm OH}$ and $Y_{{\rm H}_3{\rm O}^+}$ since these species are generated solely in water. 

On the low $R_{\rm C}$ side, the yields obtained for the two values of the coupling strength are different. This difference reflects the difference in the electron cascade in the composite system. In the case of a larger electron-phonon interaction in silica, an electron has more chance to thermalize and to recombine in silica, so that the electron yield in water is lower.

\subsection{Porosity} 
The porosity $p$ has a significant influence on the yields. In this section, we compare the yields obtained for two values of the porosity: $p=0.5$, discussed above, and $p=0.7$. For both OH and H$_3$O$^+$, the effect of changing the porosity follows closely the change of the fraction $f_{{\rm H}_2{\rm O}}$ with the porosity, and both yields can be obtained by a simple scaling of the corresponding fractions. Using Eq.~\eqref{eq4}, we have $f_{{\rm H}_2{\rm O}}=0.40$ for $p=0.5$ and $f_{{\rm H}_2{\rm O}}=0.61$ for $p=0.7$, and we obtain a scaling factor $s\approx 1.52$, which corresponds to the ratio of the yields at $p=0.7$ over those at $p=0.5$, whatever the value of $R_{\rm C}$. 

\begin{figure}
	\centering
		\includegraphics[width=0.50\textwidth]{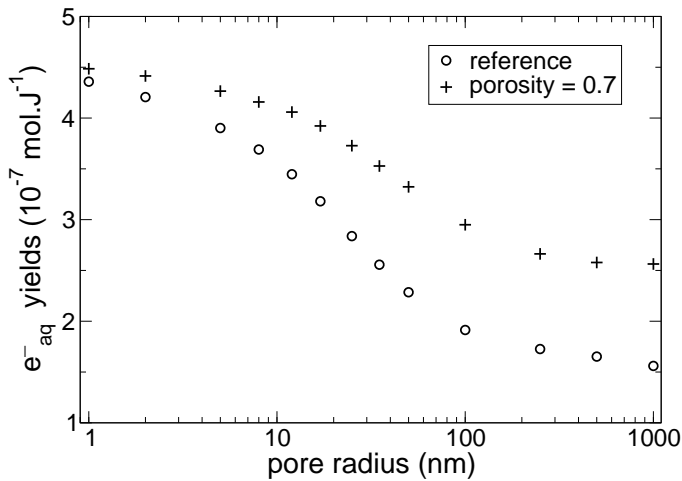}
	\caption{Yields of aqueous electrons, $e_{\rm aq}^-$, for two porosities: $p=0.5$ (reference, circles) and $p=0.7$ (crosses).}
	\label{fig7}
\end{figure} 

For the electron yield, $Y_{{\rm e}_{\rm aq}^-}$, presented in Fig. \ref{fig7} the situation
is different. From a qualitative point of view, the large increase observed as the pore radius $R_{\rm C}$ decreases is observed in both cases, and the analysis made for the $p=0.5$ yields remains valid for different porosities. The difference reflects the peculiarity of electron collection in water. For large pores, surface crossing is unimportant and the ratio of the electron yields simply reflects the above-defined scaling factor $s\approx 1.52$, like for the other yields. 

On the contrary, for small values of $R_{\rm C}$, the collection of electrons in water becomes very efficient. For porosities considered here, for which the volume of water is comparable to
the volume of silica, the electron yield is weakly sensitive to the porosity. In this limit, it is customary to consider the medium as an ideal mixture subject to an average dose delivery, identical in both water and silica. In such a case, it is possible to express the yield as a function of the mass fraction of water in the sample, $\phi_{{\rm H}_2{\rm O}}$, the yields of electrons in pure water, $\tilde{Y}_{{\rm H}_2{\rm O}}$, and the yield of electron-hole pairs in pure silica $\tilde{Y}_{{\rm SiO}_2}$: 

\begin{equation}\label{eq5}
Y = \tilde{Y}_{{\rm H}_2{\rm O}}\phi_{{\rm H}_2{\rm O}} + g_0\tilde{Y}_{{\rm SiO}_2} (1-\phi_{{\rm H}_2{\rm O}})
\end{equation} 

\noindent where $g_0$ represents the efficiency of electron transfer from silica to water. In the case of an ideally complete transfer, $g_0=1$ and, since $\tilde{Y}_{{\rm H}_2{\rm O}} < \tilde{Y}_{{\rm SiO}_2}$, $Y$ would be a decreasing function of $\phi_{{\rm H}_2{\rm O}}$. This is not the case in our simulation, as it can be observed in Fig. \ref{fig7}. Indeed, even for small silica thicknesses considered here, $g_0$ depends on $R_{\rm C}$ and on the porosity, or equivalently on $\phi_{{\rm H}_2{\rm O}}$. At $R_{\rm C}=1$ nm, we obtain $g_0=0.89$ for a porosity of 0.5 ($\phi_{{\rm H}_2{\rm O}}=0.307$) and $g_0=0.97$ for a porosity of 0.7 ($\phi_{{\rm H}_2{\rm O}}=0.509$).

\subsection{Segregation} 
\begin{figure}
	\centering
		\includegraphics[width=0.45\textwidth]{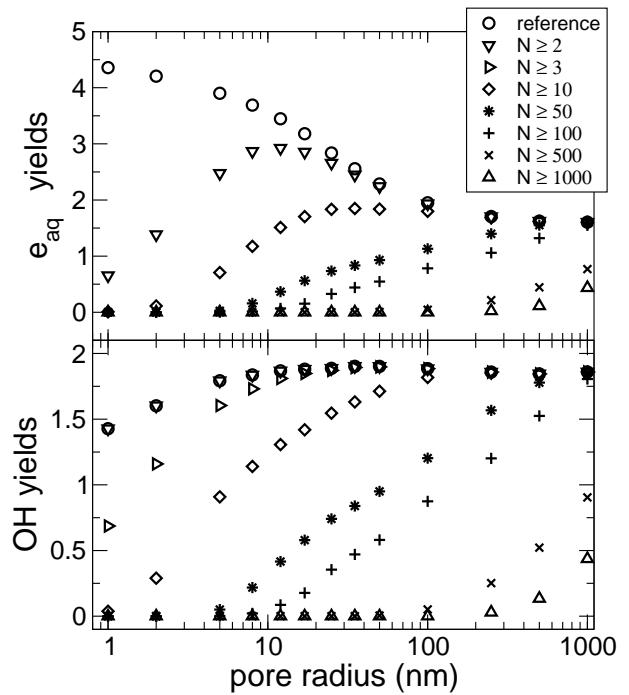}
	\caption{Segregation effect on electron (upper panel) and on hydroxyl (lower panel) yields, as a function of pore radius $R_{\rm C}$. The reference curve corresponds to the calculations of the yields including all pores, i.e., pores containing a number of species $N \ge 1$. The yields are expresses in units of 10$^{-7}$ mol/J.}
	\label{fig8}
\end{figure}

Our simulation offers an interesting qualitative picture of the distribution of radicals in pores. It is straightforward to analyze the number of radicals per pore as a function of the pore radius. This distribution can greatly influence the chemical reactions between radicals that take place when these radicals start to diffuse in pores. Such a distribution gives insight into the segregation effect, i.e., the ability of the composite system to isolate a limited number of species from the other ones. We have thus plotted in Fig. \ref{fig8} the yields obtained by taking into account \emph{only} the pores containing a number $N$ of chemical species of any kind, for which $N$ is larger than a threshold value $N_{\rm T}$. The reference curve was obtained for a threshold value $N_{\rm T}=1$, i.e., without any constraint on the number of species in a pore. 

For solvated electrons, a first and obvious observation is that, as $N_{\rm T}$ increases, the yields for a given $R_{\rm C}$ become smaller. However, as the pores become larger for a given value of $N_{\rm T}$, the yields tend to become somewhat less sensitive to the number of pores excluded from their computation, and the related curves converge toward the reference curve. The difference between the cases $N_{\rm T}=1$ (reference) and $N_{\rm T}=2$ is remarkable. It
corresponds to the contribution of singly isolated species, having thus no partner to react with. These singly isolated species are exclusively electrons, because all the other products of water radiolysis result from the fragmentation of the water molecule, giving at least two species. For very small pore sizes, the corresponding yield represents more than one-half of the total yield. Therefore, in addition to confinement, segregation plays a significant role in the radiolysis of water when pores are very small. This effect disappears progressively for $R_{\rm C} > 8$ nm. Nevertheless, for this radius, only 30\% of the pores contain more than 10 chemical species, and the chemical reactions can be quite modified. When increasing the radius $R_{\rm C}$, the number of species in a pore becomes progressively comparable to the number of species in spurs, and the distribution of the species becomes comparable to the distribution in bulk water.

For OH radicals, the yield depends on $N_{\rm T}$ and $R_{\rm C}$ in the same fashion as for the electron yields. However, the curves obtained for $N_{\rm T}=1$ (reference) and $N_{\rm T}=2$ overlap perfectly. Indeed, a single OH molecule cannot be isolated in a pore, because OH
is a dissociation product of an excited or ionized water molecule, which is always associated with the formation of a hydrogen atom or a proton. Below $R_{\rm C}=8$ nm, most of the pores contain only a few radicals, and the production of molecular species will probably be affected during the chemical stage.

\section{Conclusion}
Our simulation offers a detailed picture of the physicochemical processes involved in the radiolysis of water when it is confined in silica, and more generally when water and silica
form a composite heterogeneous system with characteristic nanometric lengths. The main effect is the collection and solvation of a large excess of electrons in water. The existence of the process is controlled exclusively by the difference between the conduction band edge of the materials. Its efficiency is controlled by the relative value of the low-energy electron mean-free
path in silica with respect to the size of the pores $R_{\rm C}$ and to the distance between pores, defined itself by the porosity $p$. 

On the basis of our simulation, we expect a similar behavior for any other material for which the conduction band edge is energetically above the conduction band edge of water. It might
thus be interesting to change the nature of the oxide in order to vary this parameter and thus to obtain a more definite proof of this proposal. Alternatively we might change the nature of the
polar liquid. Moreover, such a study could give new insight into the actual position of the conduction band edge with respect to vacuum. Conversely, for metals for which the conduction
band edge is well below the solvation energy of electron in water, we expect the opposite behavior.

Our simulation supports and refines the conclusion reported by Schatz and co-workers \cite{ref3}. The low-energy subexcitation electrons produced in silica are likely to cross the interface to
be trapped in water when their diffusion length becomes comparable to the particle size. For sufficiently small silica particle size, i.e., below 10 nm, this effect becomes extremely
efficient. This large amount of excess electrons will, of course, diffuse and react with all the products of water radiolysis. Nevertheless, it is clear that such a large excess of solvated
electrons in the picosecond range is likely to result in an excess of electrons in the nanosecond range as observed experimentally \cite{ref3}. 

The yields of the other main species, OH and H$_3$O$^+$, are much less affected by the presence of silica surrounding water. However, in the case of ideally perfect confinement, a pronounced
segregation effect is observed for small radii. Since a part of the species have no partner to react with, some specific effects may appear during the chemical stage. The analysis of
the chemical stage will be the object of future work.

\begin{acknowledgments}
This study is part of the RADICO project funded by the Agence Nationale de la Recherche, contract number ANR-07-BLAN-0358-01.
\end{acknowledgments} 

\appendix 

\section{Electron interaction cross sections}

\subsection{Ionization}

\begin{widetext}

To compute the ionization cross sections, we follow our previous work on water \cite{ref10}. In the case of silica, we consider 5 core and 10 valence energy levels. The core energies are taken as atomic orbital energies from Hartree-Fock calculation \cite{ref33}, as given in Table \ref{tab2}. The valence energy levels are fitted by a linear combination of Gaussian functions to reproduce the valence density of state (DOS) $f_{\rm V}(B)$ \cite{ref15}: 

\begin{equation}
f_{\rm V}(B) = \sum_{i=1}^{10} c_i g_i(B)
\end{equation}

\noindent with $B$ being the binding energy and where $g_i(B) = \sqrt{\eta_i/\pi} \exp\left[-\eta_i(B-B_i)^2\right]$. The coefficients $c_i$ are normalized so that $\sum_{i=1}^{10}c_i = N_{\rm V}$ , where $N_{\rm V} = 12$ is the number of valence electrons for one SiO$_2$ unit. The parameters $B_i$, $c_i$, and $\eta_i$ are given in Table \ref{tab2}.

To determine the cross sections, we also need the conduction DOS, $f_{\rm C}(K)$. For silica, the ratio of $f_{\rm C}(K)$ over an ideally parabolic DOS, which we assume to correctly describe an isolated atom, is given by the ratio of the effective mass $m^{\star}$ to the mass $m$ of a free electron: 

\begin{equation}
f_{\rm C}(K) \propto \left(\frac{m^{\star}}{m}\right)^{3/2} K^{1/2} = s_{\rm C}(K) K^{1/2}
\end{equation}

For silica, the energy dependence of the effective mass is taken from the work of Fischetti and coworkers \cite{ref26}. For water we disregard any modification of the conduction DOS. For the sake of consistency with the work of Kim and Rudd \cite{ref21}, $s_{\rm C}(K)$ is normalized so that the sum rule is satisfied for each energy level $i$:

\begin{equation}
Q_i \int_0^{\infty} \frac{s_{\rm C}(K)}{(K/B_i + 1)^2}~{\rm d}K = 1
\end{equation}

\noindent We set $Q_i s_{\rm C}(K) = s_{{\rm C},i}(K)$. With the above definitions, the differential cross section for energy loss $W = K + B$, for an electron of velocity $v$ and for level $i$ with an energy $B$ distributed around $B_i$ is given by:

\begin{equation}
\frac{{\rm d}\sigma_i}{{\rm d}W} = \int s_{{\rm C},i}(K)c_i g_i(B) \frac{{\rm d}\sigma}{{\rm d}W}{\rm d}B
\end{equation}

The ionization cross section for a binding energy $B$, thus reads \cite{ref21}:

\begin{equation}\label{crsec}
\frac{{\rm d}\sigma}{{\rm d}W} = \frac{2\pi}{B^3(t+a+1)}~\times \sum_{n=1}^3 \left((w+1)^{-n}+(t-w)^{-n}\right)F_n(t),
\end{equation}

\noindent where

\begin{equation}
F_1 = -\frac{F_2}{t+1}, ~~~~F_2 = \frac{2-q}{t+1}, ~~~~F_3 = \frac{q\ln(t)}{t+1}.
\end{equation}

\noindent The cross section in Eq.~\eqref{crsec} is expressed in dimensionless variables: $t = mv^2/2B$, $w = W/B$, $k = K/B = w - 1$, and $a = A/B$. In the present work, we used $q=1$. 

Following the proposal of Kim and Rudd \cite{ref21}, the value for $A$ should be the orbital kinetic energy for a given shell. However, these authors used a set of values twice as large, i.e. $A = 2 \langle p2/2m\rangle$, to compute the ionization cross sections of water molecules. This choice is not consistent with the parameters used for rare gas atoms, but it gives good results for water molecules. We therefore decided to use this scaling for the valence states of liquid water. Since little is known regarding the ionization cross section of molecules containing Si atoms, we adopted the same scaling for the valence states of SiO2. For the core states, we do not scale this parameter and use $A = \langle p^2/2m \rangle$. The values used in the simulation are given
in tables \ref{tab2} and \ref{tab3}. They are determined for each energy level from the corresponding atomic levels in the Hartree-Fock approximation \cite{ref33}. The effect of scaling $A$ by a factor 2 is to reduce by approximately 30 \% the cross section for energies below 100 eV. In order to check the sensitivity of our results to one particular choice of $A$, we computed the yield $G$ of electron-hole pairs for both sets of parameters. For silica, we obtain $G = 5.4 \times 10^{−7}$ mol/J without scaling and $G = 5.2 \times 10^{−7}$ mol/J with the factor 2 scaling. The consequences for electron transfer from silica to water, as well as for the segregation of radicals is marginal. Our parameterization reproduces the recommended stopping power values \cite{ref34} within 2\% in the energy range 10 - 50 keV for both H$_2$O and SiO$_2$.

\begin{table}[!rh]
\centering
\caption{Parameters used to calculate the single ionization cross sections in $\mbox{H}_2\mbox{O}$. Note that $A=2\langle p2/2m\rangle$ for all valence levels. All values are given in atomic unit of energy: 1 hartree = 27.21 eV.}
\label{tab2}
\begin{tabular}{llllc}
\hline\hline
~&~&~&~&~\\
$c_i$& 		$B_i$(hartree)&	$\eta^{-1/2}$(hartree)& 	$A$(hartree)&	Molecular states\\
~&~&~&~&~\\
core states&~&~&~&~\\
\hline\\
2.0&					  19.77& 0.02& 					 29.26&	$\mbox{H}_2\mbox{O}$(1$a$1)\\
2.0&	\phantom{1}1.19& 0.02& \phantom{2}2.61&	$\mbox{H}_2\mbox{O}$(2$a$1)\\
&~&~&~&~\\
valence states&~&~&~&~\\
\hline\\
2.0& 0.61& 0.029&	3.58&	$\mbox{H}_2\mbox{O}$(1$b$2)\\
2.0& 0.54& 0.068&	4.35&	$\mbox{H}_2\mbox{O}$(3$a$1)\\
2.0& 0.44& 0.068&	4.52&	$\mbox{H}_2\mbox{O}$(1$b$1)\\
\hline\hline
\end{tabular}
\end{table}

\begin{table}[!rh]
\centering
\caption{Parameters used to calculate the single ionization cross sections in $a\mbox{-SiO}_2$. Note that $A=2\langle p2/2m\rangle$ for all valence levels. The average orbital kinetic energy
$\langle p2/2m\rangle$ is computed from the table of Clementi and Roetti \cite{ref33}. All values are given in atomic unit of energy: 1 hartree = 27.21 eV.}
\label{tab3}
\begin{tabular}{llllc}
\hline\hline
~&~&~&~&~\\
$c_i$& 		$B_i$(hartree)&	$\eta^{-1/2}$(hartree)& 	$A$(hartree)&	Atomic levels\\
~&~&~&~&~\\
core states&~&~&~&~\\
\hline\\
2.0&						 67.91&	0.02&					  92.17&	Si(1$s$)\\
2.0&						 19.77& 0.02&					  29.26&	O(1$s$)\\
2.0&	 \phantom{1}5.81&	0.02&						13.81&	Si(2$s$)\\
6.0&	 \phantom{1}3.97&	0.02&			      12.14&	Si(2$p$)\\
2.0&   \phantom{1}1.21&	0.02&	\phantom{1}3.16&	O(2$s$)\\
&~&~&~&~\\
valence states&~&~&~&~\\
\hline\\
1.19& 0.68&	0.029&		2.70& Si(3s)\\
1.19&	0.59&	0.029&		2.70& Si(3s)\\
0.75& 0.54&	0.004&		2.20&	Si(3p)\\
0.99& 0.50&	0.007&		4.96&	O(2p)\\
2.61& 0.42&	0.029&		2.20&	Si(3p)\\
1.86& 0.40&	0.008&		2.20&	Si(3p)\\
0.60& 0.38&	0.008&		2.20&	Si(3p)\\
0.67& 0.36&	0.008&		4.96&	O(2p)\\
0.05& 0.35&	0.003&		4.96&	O(2p)\\
2.09& 0.34&	0.016&		4.96&	O(2p)\\
\hline\hline
\end{tabular}
\end{table}

\subsection{Excitation} 
For an electron with velocity $v$, the differential cross section for energy loss $W$ by excitation reads: 

\begin{equation}
\frac{{\rm d}\sigma}{{\rm d}W} = \frac{\pi}{T+W+A}~\frac{f(W)}{W}~\ln \left(\frac{q_{\rm max}}{q_{\rm min}}\right)
\end{equation}

\noindent where $T = mv^2/2$, and: 

\begin{equation}
q_{\rm min} = 2t -1 - 2\sqrt{t(t-1)}
\end{equation}

\begin{eqnarray}
q_{\rm max} = 2t - 1 +2\sqrt{t(t-1)} ~~~~ & \mbox{if} & ~~~~ q_{\rm max} < 1 \\
q_{\rm max} = 1 ~~~~ & \mbox{if} & ~~~~ q_{\rm max} \ge 1
\end{eqnarray} 

\noindent where $t = T/W$. For exciton creation in silica, $A$ = 0.33 a.u. (atomic units) or $A$ = 9.0 eV, and the distribution of oscillator strength $f(W)$ assumes a Gaussian shape: $f(W) = f_0 \sqrt{\eta/\pi} \exp\left[-\eta(W - W_0)^2\right]$, with $f_0 = 0.237$, $\eta^{-1/2} = 0.06$ a.u.  and $W_0 = 0.32$ a.u. = 8.7 eV.

\end{widetext}

\end{document}